\documentclass[fleqn,usenatbib]{mnras}

\usepackage{newtxtext,newtxmath}

\usepackage[T1]{fontenc}

\DeclareRobustCommand{\VAN}[3]{#2}
\let\VANthebibliography\thebibliography
\def\thebibliography{\DeclareRobustCommand{\VAN}[3]{##3}\VANthebibliography}

\usepackage{graphicx}	
\usepackage{amsmath}	
\usepackage{amssymb}	
\usepackage{booktabs}
\usepackage{xcolor}
\usepackage{ulem}
\usepackage{comment}

\newcommand{\Rcrit}{\ifmmode{R_{\mathrm{crit}}}\else$R_{\mathrm{crit}}$\fi}
\newcommand{\rcl}{\ifmmode{r_{\mathrm{cl}}}\else$r_{\mathrm{cl}}$\fi}
\newcommand{\lsh}{\ifmmode{l_{\mathrm{sh}}}\else$l_{\mathrm{sh}}$\fi}
\newcommand{\Tcl}{\ifmmode{T_{\mathrm{cl}}}\else$T_{\mathrm{cl}}$\fi}
\newcommand{\Tfl}{\ifmmode{T_{\mathrm{fl}}}\else$T_{\mathrm{fl}}$\fi}
\newcommand{\Nclumps}{\ifmmode{N_{\mathrm{clumps}}}\else$N_{\mathrm{clumps}}$\fi}
\newcommand{\ncl}{\ifmmode{n_{\mathrm{cl}}}\else$n_{\mathrm{cl}}$\fi}
\newcommand{\nfl}{\ifmmode{n_{\mathrm{fl}}}\else$n_{\mathrm{fl}}$\fi}
\newcommand{\Tmix}{\ifmmode{T_{\mathrm{mix}}}\else$T_{\mathrm{mix}}$\fi}
\newcommand{\Twind}{\ifmmode{T_{\mathrm{wind}}}\else$T_{\mathrm{wind}}$\fi}
\newcommand{\Thot}{\ifmmode{T_{\mathrm{hot}}}\else$T_{\mathrm{hot}}$\fi}
\newcommand{\Tdest}{\ifmmode{T_{\mathrm{dest}}}\else$T_{\mathrm{dest}}$\fi}
\newcommand{\tsc}{\ifmmode{t_{\mathrm{sc}}}\else$t_{\mathrm{sc}}$\fi}
\newcommand{\tcc}{\ifmmode{t_{\mathrm{cc}}}\else$t_{\mathrm{cc}}$\fi}
\newcommand{\dcell}{\ifmmode{d_{\mathrm{cell}}}\else$d_{\mathrm{cell}}$\fi}
\newcommand{\chiF}{\ifmmode{\chi_{\mathrm{F}}}\else$\chi_{\mathrm{F}}$\fi}
\newcommand{\chiI}{\ifmmode{\chi_{\mathrm{I}}}\else$\chi_{\mathrm{I}}$\fi}

\newcommand{\tminmix}{\ifmmode{t_{\mathrm{cool,minmix}}}\else$t_{\mathrm{cool,minmix}}$\fi}
\newcommand{\tmax}{\ifmmode{t_{\mathrm{cool,max}}}\else$t_{\mathrm{cool,max}}$\fi}
\newcommand{\tmixCC}{\ifmmode{t_{\mathrm{dest,mix}}}\else$t_{\mathrm{dest,mix}}$\fi}
\newcommand{\tmixCool}{\ifmmode{t_{\mathrm{cool,mix}}}\else$t_{\mathrm{cool,mix}}$\fi}
\newcommand{\tcool}{\ifmmode{t_{\mathrm{cool}}}\else$t_{\mathrm{cool}}$\fi}

\newcommand{\cc}{cm$^{-3}$}

\newcommand{\Mach}{\ifmmode{\mathcal{M}}\else$\mathcal{M}$\fi}

\newcommand{\rattering}{splintering}

\title[Molecular Shattering]{Molecular Shattering}

\author[Farber \& Gronke]{
Ryan J. Farber,$^{1}$\thanks{E-mail: rjfarber@umich.edu}
and Max Gronke,$^{1}$
\\
$^{1}$Max Planck Institute for Astrophysics, Karl-Schwarzschild-Str. 1, D-857481 Garching, Germany \\
}

\date{Draft from \today}

\pubyear{2022}

\begin{document}
\label{firstpage}
\pagerange{\pageref{firstpage}--\pageref{lastpage}}
\maketitle

\begin{abstract}
Recent observations suggest galaxies may ubiquitously host a molecular component to their multiphase circumgalactic medium (CGM).
However, the structure and kinematics of the molecular CGM remains understudied theoretically and largely unconstrained observationally.
Recent work suggests molecular gas clouds with efficient cooling survive acceleration in hot winds similar to atomic clouds. Yet the pressure-driven fragmentation of molecular clouds when subjected to external shocks or undergoing cooling remains unstudied.
We perform radiative, inviscid hydrodynamics simulations of clouds perturbed out of pressure equilibrium to explore the process of hydrodynamic fragmentation to molecular temperatures.
We find molecular clouds larger than a critical size can shatter into a mist of tiny droplets, with the critical size deviating significantly from the atomic case. 
We find that cold clouds shatter only if the sound crossing time exceeds the local maximum of the cooling time $\sim$8000\,K.
Moreover, we find evidence for a universal mechanism to `shatter' cold clouds into a `mist' of tiny droplets as a result of rotational fragmentation -- a process we dub `\rattering.'
Our results have implications for resolving the molecular phase of the CGM in observations and cosmological simulations.
\end{abstract}



\section{Introduction}
\label{sec:intro}
\vspace{-0.1cm}
The recent development of state-of-the-art submillimeter arrays (e.g., ALMA) and integral field spectrograph instruments is beginning to shed light on the evolution of the molecular phase in the galactic baryon cycle \citep[][]{tumlinson2017circumgalactic}. The cold molecular phase is evidently ubiquitous, detected in galactic outflows and the circumgalactic medium (CGM) both from within the Local Group \citep[][]{Di2019,Di2020,su2021molecular}, nearby starbursts (M82 \citealt{strickland2009supernova} , NGC 253 \citealt{strickland2000chandra,strickland2002chandra}), close AGN (e.g., Cen A; \citealt{Charmandis2000}) and at high redshifts \citep[][]{walter2004resolved,Cicone2014,schumacher2012gas,weiss2012variations,ginolfi2017molecular,vayner2017galactic,stacey2022red,stacey2022luck}. 

The hot ionized phase of galactic winds has been well motivated theoretically and well studied observationally for decades, \citep[][]{Chevalier1985,Martin2005,Strickland2007}, yet a proper physical understanding for multiphase winds has only recently begun to emerge \citep[][]{HuangS2020,Veilleux2020,Schneider2020,fielding2022structure}. The observed coexistence of a cold molecular phase is theoretically perplexing. Cold gas should be disrupted due to hydrodynamical instabilities \citep[][]{Klein1994,zhang2017entrainment} and purportedly the CGM does not require a molecular phase to account for all missing baryons \citep[][]{werk2014cos}. Yet the molecular phase tends to dominate the gas mass in the interstellar medium \citep[][]{Draine2011} and its recycling plays a crucial role in resolving the `gas depletion time problem' in explaining the longevity of galactic star formation \citep[][]{bigiel2011constant}.

Moreover, observations detect massive reservoirs of outflowing dense, molecular gas \citep[][]{Cicone2014,cicone2021super,munoz2022apex}. Whether this gas survives the acceleration process or is restricted to small heights beyond the galactic midplane depends crucially on the size distribution of molecular clouds embedded in those hot winds \citep[][]{farber2022survival}.

Previous work has considered the resulting cloud size distribution of warm $\sim$10$^4$\,K clouds undergoing isochoric thermal instability due to rapid cooling \citep[][]{McCourt2018,Waters2019,Gronke2020Misty,Das2021}. \citet[][]{McCourt2018} suggested clouds hierarchically `shatter' via a process analogous to Jeans instability but driven by an excess external pressure rather than gravity. The characteristic size of `shattered' fragments \lsh\ $\sim$ min($c_s$ $t_{\rm cool}$) $\sim$ 0.1 pc (\cc/$n$), where $n$ is the gas number density, suggests the CGM should be a mist of tiny droplets, explaining the large area covering fraction yet small volume of observed systems \citep[reviewed in][]{McCourt2018}. 

\citet[][]{Waters2019} performed 1D models of long wavelength entropy modes and pointed out clouds might fragment due to a rather different mechanism they term `splattering.' Rapidly cooling clouds encounter a `cooling wall' at $\sim$8000\,K (when Hydrogen becomes fully neutral) causing a reversal of their motion (from contraction to explosion). The rebounding clouds' velocity can exceed their sound speed, hence fragmenting when their pressure exceeds the ambient. 
\citet[][]{Gronke2020Misty} performed three dimensional hydrodynamic simulations of clouds cooling from $\sim$10$^6$\,K to $10^4$\,K, finding fragmentation occurred during a `rebound' after clouds cooled to their temperature floor. 

In this work, we extend these previous studies by considering cooling down to molecular temperatures $\sim$400\,K. We find evidence for a two-stage fragmentation process, including a new rotational mechanism for shattering,\footnote{A movie of the density evolution is available at \url{https://tinyurl.com/molecularShattering}, and a movie of the temperature evolution at higher resolution is available at \url{https://tinyurl.com/molecularShattering2}.} which we term `\rattering.' Our results have implications for the size distribution of molecular gas in galactic halos.

\vspace{-0.5cm}
\section{Methods}
\label{sec:Methods}
\vspace{-0.1cm}
We performed our simulations with a modified version of FLASH4.2.2 \citep[][]{Fryxell2000,Dubey2008,Farber2018,farber2022survival}, utilizing the unsplit staggered mesh solver \citep[][]{Lee2009,Lee2013} to solve the compressible, Eulerian fluid equations with radiative cooling included as a sink term. We utilize the second-order MUSCL-Hancock predictor-corrector method with the HLLC Riemann solver and \texttt{mc} slope limiter. 

To model optically thin radiative cooling we use the \citet{townsend2009} exact integration scheme to improve the stability of our solutions and to model the cooling process to machine precision accuracy.\footnote{Rather, the principal source of error is relegated to interpolation from fits to the cooling curve.} We apply a piecewise power law fit to the \citet{Sutherland1993} cooling curve down to 10$^4$\,K, extended down to 300\,K with the cooling curve of \citet{Dalgarno1972}. Note that we turn off radiative cooling above 0.6 \Thot\ to roughly model the effect of radiative heating of the tenuous ambient medium (e.g., due to the metagalactic radiation field, \citealt{haardt2012radiative}; \citealt{faucher2020cosmic}).

In each of our simulations we initialize four overlapping spherical clouds at a thermally unstable mean temperature \Tcl\ = 4 $\times 10^5$\,K and mean density \ncl\ = 10 \cc\ in hydrostatic equilibrium with a hot ambient medium \Thot\ = $4 \times 10^7$\,K and an initial density contrast \chiI\ = 100. One cloud was placed at domain center with the other three offset at most one cloud radius \rcl\ per dimension to introduce asymmetry into our initial conditions \citep{Gronke2020Misty}\footnote{Asymmetry mitigates the carbuncle instability and is more representative of observed filamentary clouds \citep{young2005spitzer}.}. We also perturb the density for cloud material by drawing from a Gaussian with ($\mu$, $\sigma$) = (1, 0.01) \ncl, truncated at 3$\sigma$. We set the temperature to maintain a fixed initial pressure of 4 $\times 10^6$ \cc\ K. 

We utilize static mesh refinement in our simulations to ensure uniform coverage at the center of the domain with an extended box at lower resolution to prevent complete outflow of material and mitigate the effect of boundary conditions on our simulation results. Specifically, we perform our simulations in a box with resolution 16 cells per \rcl\ in the innermost (8 \rcl)$^3$, outside of which the resolution degrades at most a factor of four to a full box size of (24 \rcl)$^3$.

We performed a suite of simulations varying the imposed temperature floor \Tfl\ from 400\,K to 4 $\times 10^5$\,K and physical scale of \rcl\ from $10^{-4}$\,pc to 10\,kpc to cover uniformly (in log-space) five decades in temperature  and ten decades in cloud size. We find these two parameters play a key role in the dynamical evolution of clouds cooling to molecular temperatures as we now show. 

\vspace{-0.5cm}
\section{Results}
\label{sec:Results}
\begin{figure}
  \begin{center}
    \leavevmode
    \includegraphics[width=\linewidth]{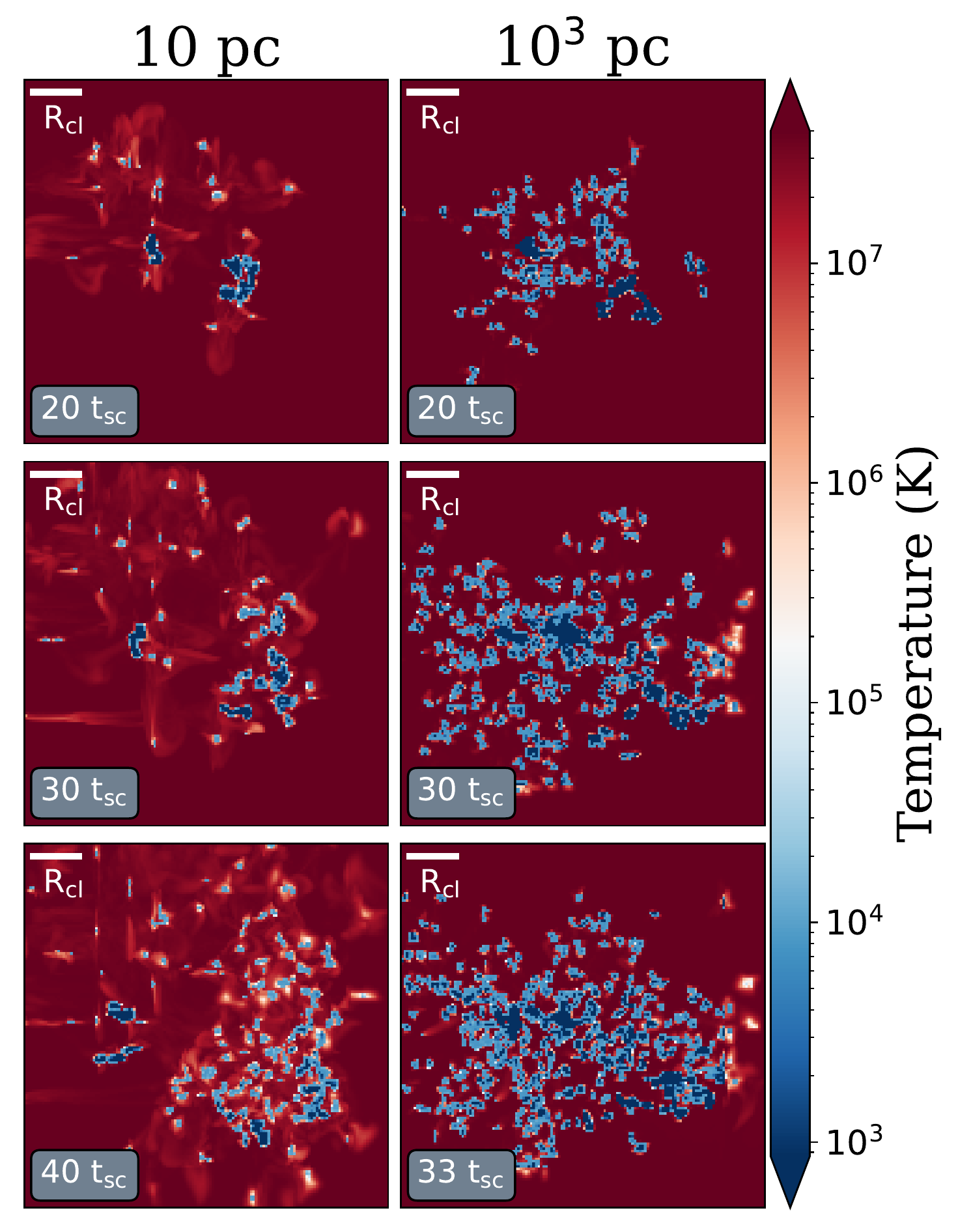}
\vspace{-0.5cm}
\caption[]{Projections of the minimum temperature along the line of sight. While most of the clumps in the large cloud case (right column) have gas close to the floor temperature, clumps in the small cloud case (left column) are largely at or above 10$^4$ K.}
\vspace{-0.5cm}
\label{fig:TemperatureProjections}
\end{center}
\end{figure}

\vspace{-0.1cm}
We investigate the dependence of shattering on (i) initial cloud size \rcl\ and (ii) floor temperature \Tfl. In Fig. \ref{fig:TemperatureProjections} we graphically display how shattering depends on \rcl\ at a fixed \Tfl\ $\sim$ 800\,K via minimum temperature projections\footnote{Note that the evolution is largely three-dimensional so many clumps are missed when looking at slice plots. Moreover, mass-weighted projections still favor the substantial shells of warm 10$^4$\,K gas at the expense of embedded cold 10$^3$\,K gas. We choose to display minimum temperature projections to specifically determine the evolution of the coldest phase.}.

Comparing the smaller \rcl\ $\sim$ 10\,pc cloud (left) to the larger 10$^3$\,pc cloud we see that the smaller cloud: (i) fragments into fewer clumps than the large cloud; (ii) takes more time to form a large number of clumps; (iii) mixes more with the hot ambient medium (white colors); (iv) only possesses $T \sim 10^4$\,K gas in a few large clumps, whereas most of the large cloud's clumps (right column) possess \Tfl\ gas.

\begin{figure}
  \begin{center}
    \leavevmode
    \includegraphics[width=\linewidth]{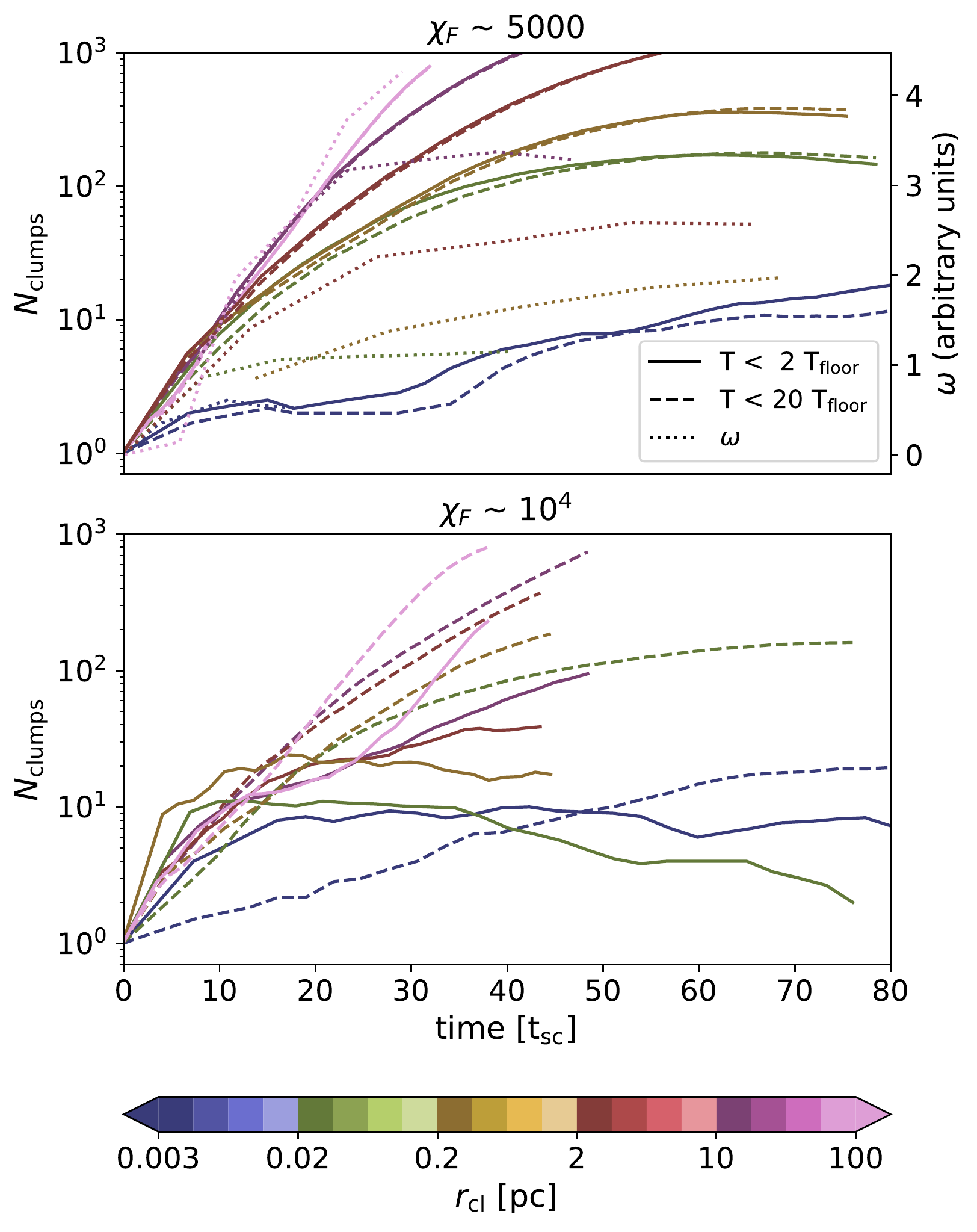}
    \vspace{-0.5cm}
\caption[]{Time evolution of the number of clumps (and angular velocity $\omega$ for the top panel) formed in a batch of our simulations. The top panel shows the \chiF\ $\sim$ $5 \times 10^3$ case and the bottom panel shows the \chiF\ $\sim$ $10^4$ case with curves colored to indicate initial cloud radius from \rcl\ $\sim$ 0.003\,pc (blue) to \rcl\ = 100\,pc (pink).}
\vspace{-0.5cm}
\label{fig:NumClumps}
\end{center}
\end{figure}

\noindent 
We next examine more closely the dependence on the relative number of clumps for different gas phases as a function of initial cloud radius.
In Fig. \ref{fig:NumClumps} we have plotted the number of clumps vs. time, as well as the median rotational velocity $\omega$ of the clumps vs. time in the upper panel (dotted lines).
We depict cloud radii as different colors from blue (small) to pink (large).
We indicate two temperature cuts: $T <$ 2 \Tfl\ as solid curves and T $<$ 20 \Tfl\ as dashed curves. 
We plot simulations with \Tfl\ $\sim$ 8000\,K in the top panel and \Tfl\ $\sim$ 4000\,K in the bottom panel. In both panels we see that smaller clouds take longer to fragment than larger clumps and form fewer clumps (as also seen in Fig. \ref{fig:TemperatureProjections}). The maximum number of clumps across phases appears to be relatively independent of \Tfl\ at fixed \rcl\ (for the values of \Tfl\ shown here).

However, while the top panel clouds show roughly the same number of clumps with the two temperature cuts (suggesting a single phase), the bottom panel clearly shows two populations of clumps exist: a larger number of warm clumps and a smaller number of cold clumps (suggesting two phases co-exist). Note the separation between the number of clumps in each phase diminishes with increasing \rcl\ for the bottom panel (similar to the large cloud in Fig. \ref{fig:TemperatureProjections} showing almost all its clumps consisting of a \Tfl\ phase while the smaller cloud had only a few clumps at \Tfl).

Note the large change in the number of clumps from $\sim$10 for \rcl\ $\sim$\,0.003\,pc to $\sim$100 for \rcl\ $\sim$ 0.02\,pc.
To determine what separates (i) small clouds which barely fragment, (ii) larger clouds with two distinct populations, and (iii) even larger clouds with roughly all clumps at \Tfl, we next compare the sound crossing time to the cooling time.

\begin{figure}
  \begin{center}
    \leavevmode
    \includegraphics[width=\linewidth]{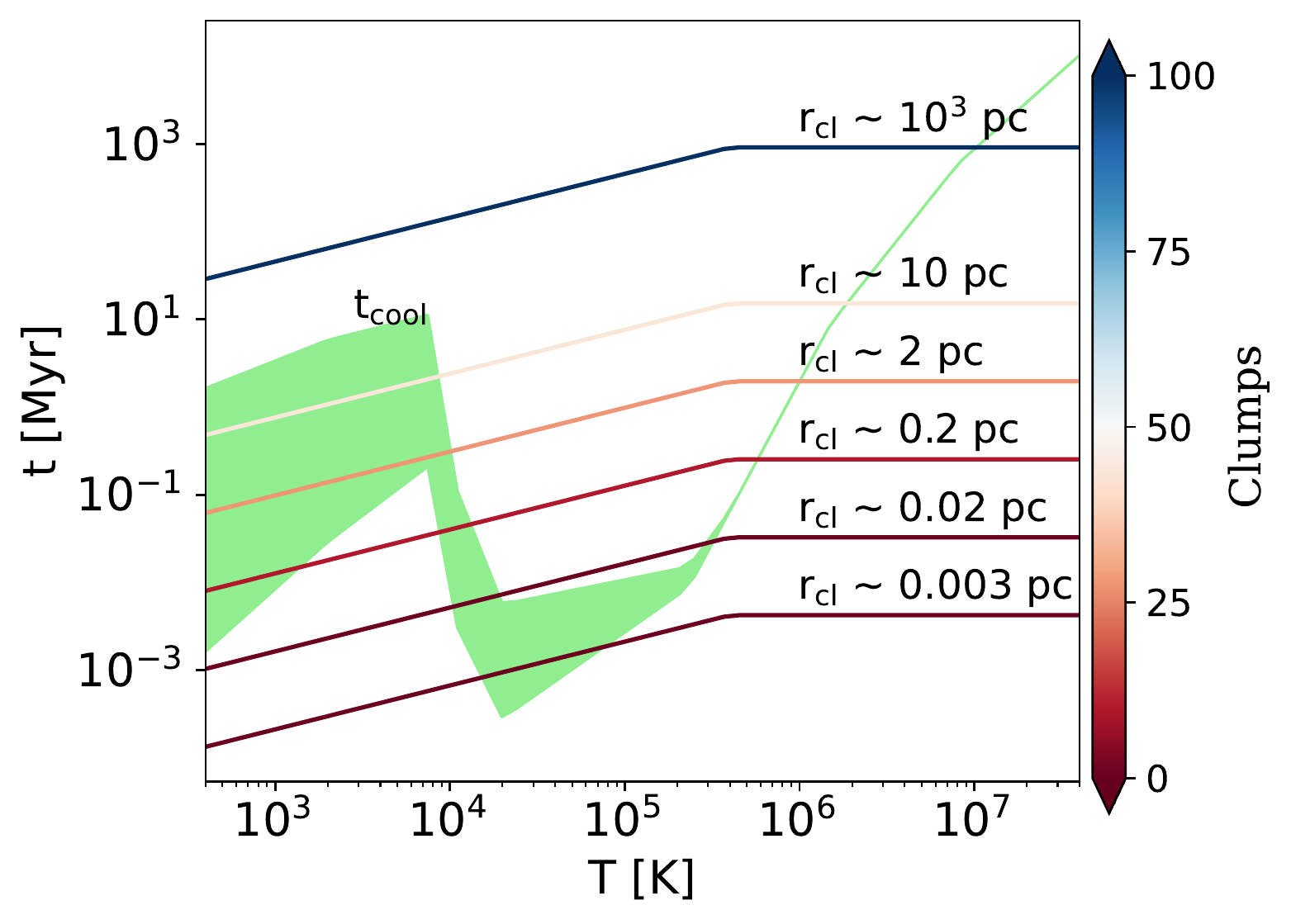}
    \vspace{-0.5cm}
\caption[]{Cooling time and sound crossing times vs. temperature. We plot the isobaric and isochoric cooling times (lower and upper region of the lightgreen band respectively). We label the sound crossing time curves with the initial cloud radius from simulations at \chiF\ $\sim$ 5 $\times 10^4$. The colors of the sound crossing time curves correspond to the maximum number of clumps formed during the simulation. Clearly the number of clumps increases as the sound crossing time exceeds the cooling time. Note the sound crossing time diminishes below \Tcl\ as the clouds contract as they cool.}
\vspace{-0.5cm}
\label{fig:Timescales}
\end{center}
\end{figure}

In Fig. \ref{fig:Timescales} the green band indicates the range in cooling times depending on whether cloud material cools isobarically (lower bound) or isochorically (upper bound).
We plot the sound crossing times for the clouds we simulated with \Tfl\ $\sim$ 800\,K with bluer colors indicating small clouds and pinker colors indicating larger clouds (as in Fig. \ref{fig:NumClumps}).
Evidently for the largest cloud case all material has a shorter cooling time than sound crossing time and therefore all clumps cool uninhibited to \Tfl, violently fragmenting thereafter.
Smaller clouds are in sonic contact to lower temperatures and re-attain sonic contact at higher temperatures, reducing the effective pressure jump small clouds experience when their cooling switches from isochoric to isobaric at $\sim$ 10$^4$\,K.
Thus only the densest cores for the \rcl\ $\sim$\ 10\,pc cloud were able to cool down to \Tfl, as we observed in Fig. \ref{fig:TemperatureProjections}.

\begin{figure}
  \begin{center}
    \leavevmode
    \includegraphics[width=\linewidth]{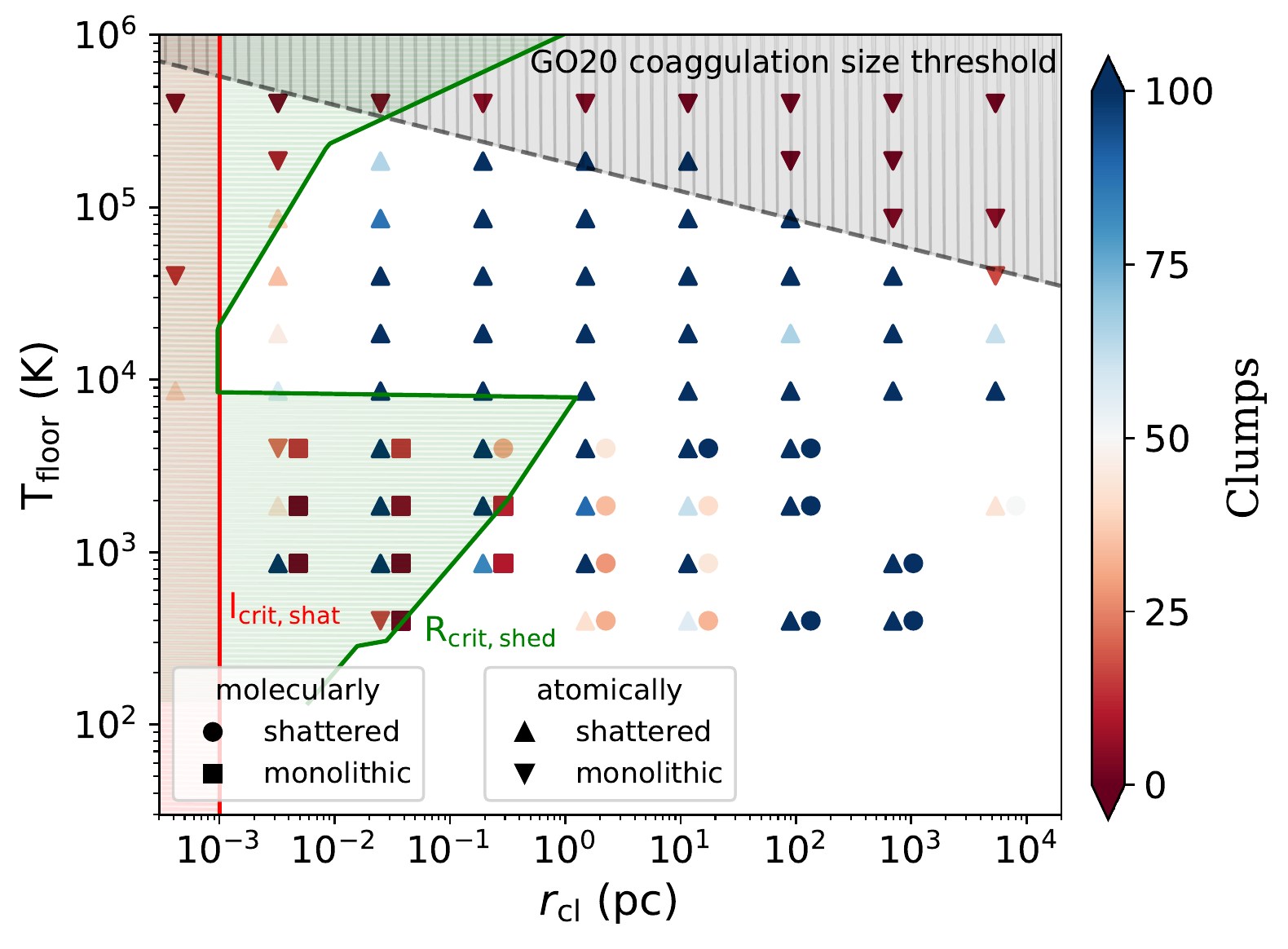}
    \vspace{-0.5cm}
\caption[]{Comparison of simulation results with critical conditions for shattering. Right-side-up (up-side-down) triangles indicate greater than (fewer than) 25 clumps with temperature $<$ 20 \Tfl\ (with characteristic temperatures of $\sim$10$^4$\,K which we refer to as atomic shattering) formed in a simulation. For simulations with \Tfl\ $<$ 8000\,K we additionally mark circles (squares) horizontally offset from the triangles to indicate molecular shattering (lack thereof) for greater than (less than) 25 clumps with temperature $<$ 2 \Tfl. We indicate the number of clumps by coloring each symbol from dark red for one clump to dark blue for 100 clumps. The coagulation size criterion of \citet{Gronke2020Misty} is included as the grey dashed curve, the atomic shattering criterion is indicated by the red vertical curve, and our predicted critical size for molecular shattering is drawn as the green curve.}
\vspace{-0.5cm}
\label{fig:Overview}
\end{center}
\end{figure}

We tested our expectation that \tcool\ $\ll$ \tsc\ for molecular fragmentation via a suite of simulations in Fig. \ref{fig:Overview}. We performed a grid of simulations with \Tfl\ $\in$ (400, 4 $\times$ 10$^5$)\,K and \rcl\ $\in$ (10$^{-4}$, 10$^4$)\,pc logarithmically spaced for even coverage and to further test the atomic shattering criteria of \citet[][]{Gronke2020Misty} in a wider parameter regime of \Tfl\ than they covered. We plot the number of clumps (from zero as red to $>100$ as blue) in two temperature cuts (T $<$ 20 \Tfl\ as right-side up / upside-down triangles and T $<$ 2 \Tfl\ as circles/squares) with the symbols indicating the final state being shattered, or remaining monolithic respectively for atomic, and molecular gas respectively (hence we only show circles/squares for \Tfl $\lesssim$ 8000\,K).

As expected, for \Tcl\ = \Tfl\ no shattering occurs.
As \Tfl\ drops more clouds atomically shatter, in agreement with the \chiF\ $\gtrsim$ 300 criterion of \citet[][]{Gronke2020Misty}.
We find remarkably good agreement between our simulations and the \citep[][]{Gronke2020Misty} coagulation size threshold $\sim$(\rcl\ / \lsh)$^{1/6}$ plotted as the grey curve.
For \Tfl\ $<$ 8000\,K we find that sufficiently small clouds fail to molecularly shatter (red squares) while they typically still atomically shatter (blue right-side-up triangles). The critical size for molecular shattering indicated by the green curve is \Rcrit\ = (\tcool\ $c_s$|$_{T_{\mathrm{loc,min}}}$ where (\tcool\ $c_s$) is evaluated at $T_{\mathrm{loc,min}}$ the temperature at which the cooling time is at a local minimum. That is, $T_{\mathrm{loc,min}} = 2 \times 10^4$\,K for 8000\,K $<$ T $<$ 2 $\times 10^4$\,K else \Tfl.
Also here, we find good agreement between our molecular shattering criterion and the results of our simulations.

\vspace{-0.7cm}
\section{Discussion}
\label{sec:Discussion}
\vspace{-0.1cm}
In this work, we extended the study of cold gas fragmentation \citep{McCourt2018,Waters2019,Gronke2020Misty,Das2021} into the colder, molecular regime. We found that $<10^4\,$K gas fragmentation does occur if\footnote{for 300\,K $<$ \Tfl $<$ 2000\,K; for 2000\,K $<$ \Tfl $<$ 8000\,K \rcl\ $>$ 300 \text{pc} (\Tfl/2000K)$^0$(
\nfl/0.1\cc)$^{-1}$ } 
\begin{equation}
r_{\rm cl} > \text{min}(c_{\rm s}t_{\rm cool})\big|_{T<8000\,K} \sim  3 \text{\,pc\,\ } T_{\rm fl,300}^{-0.5} \ n_{\rm fl,100}^{-1}
    \label{eq:molshattering_crit} 
\end{equation}
where $T_{\rm fl,300}\equiv \Tfl/300\,{\rm K}$ and $n_{\rm fl,100}\equiv \nfl/100 {\rm cm}^{-3}$.
This condition is akin to the \citet{McCourt2018} criterion for atomic shattering $r > {\rm min}(c_{\rm s}t_{\rm cool})$. However, the atomic shattering criterion is evaluated for $T > 8000$\,K whereas the minimum in Eq.~\eqref{eq:molshattering_crit} is the local minimum beyond the `bump' in the cooling time curve at $T\lesssim 8000\,$K (cf. Fig.~\ref{fig:Timescales}).
In other words, the proposed criterion for molecular shattering compares the sound crossing time to the minimum cooling time $\lesssim 8000\,$K; i.e., the timescales when the cloud can gain and lose pressure equilibrium, respectively. It follows, hence, the logic of the original `shattering' idea by \citet{McCourt2018}.

However, we found the number of clumps \textit{and} rotational velocity increased in tandem (cf. Fig.~\ref{fig:NumClumps}), suggesting centrifugal forces play a role in fragmentation, which appears evident in animations of the morphological evolution. During their trajectory in the hot medium, the droplets experience shear forces which can strip off and mix the cold gas. In \citet{farber2022survival}, we discussed the survival of molecular gas in a `cloud crushing' setup and found that the coldest gas can survive if $t_{\rm cool,max}/t_{\rm cc} < 1$ where $t_{\rm cool,max}$ is the maximum cooling time (i.e., for our employed cooling curve at $\sim 8000\,$K) and $t_{\rm cc}\sim \chi^{1/2} r_{\rm cl} / v_{\rm wind}$ is the `cloud crushing' timescale it takes a cold cloud of size $r_{\rm cl}$ to be mixed into the hot wind of velocity $v_{\rm wind}$ (in the adiabatic case). Indeed, we notice small clumps that are ejected at large relative velocity are ablated by the hot ambient medium (as seen by the white, mixing material in Fig. \ref{fig:TemperatureProjections}).

Comparing the criteria for `molecular shattering' put forward in this paper and the survival criterion for cold gas from \citet{farber2022survival} adapted to the droplets launched, we find:
\begin{align}
    \frac{t_{\rm cool,min}}{t_{\rm sc}}\frac{t_{\rm cc}}{t_{\rm cool,max}} =& \frac{t_{\rm cool,min}}{t_{\rm cool,max}}\chi^{1/2}\frac{r_{\rm d}}{r_{\rm cl}}\frac{c_{\rm s,fl}}{v_{\rm d}}\\
    \sim & \left(\frac{\chi}{10^4}\right)^{-3/2} \left(\frac{T}{T_{0.3}}\right)^{\alpha}\frac{c_{\rm s,fl}}{v_{\rm d}}\frac{r_{\rm d}}{r_{\rm cl}}
    \label{eq:crit_ratio}
\end{align}
where we normalized the quantities to $T_{\rm floor} = T_{0.3}\equiv 300\,$K, and $\Lambda(T)/\Lambda(T_{0.3})\sim (T/T_{0.3})^{\alpha}$ describes the cooling curve at these low temperatures. Note that since $t_{\rm cool}(T_{0.3})/t_{\rm cool,max}\sim 10^{-2}$ the prefactors cancel with the $\chi/10^4$ term. Furthermore, since the cloud pulsated on a timescale of $\sim t_{\rm sc}$, the `launching' velocity $v_{\rm d}\sim c_{\rm s,fl}$. While the `typical' droplet radius $r_{\rm d}$ is difficult to assess, it is reasonable to assume that the cloud breaks up into several fragments, thus, $r_{\rm cl}/r_{\rm d}\sim $a few. 
This implies that for $T_{\rm floor}\sim 300\,$K, for $\chi \sim 10^4$ the two criteria actually agree.
Furthermore, the cooling curve employed here has a slope of $\alpha \sim 1.5$ at $T\lesssim 8000\,$K. For the temperature range considered we have $t_{\rm cool,min}/t_{\rm sc}\sim (\chi/10^4)^{-1} t_{\rm cool,max}/t_{\rm cc}$, hence the two criteria actually agree for the entire temperature range considered.

While the `shattering' criteria are as described by \citet{McCourt2018}, it is noteworthy that the dynamics of fragmentation is vastly different. Both in the atomic case \citep[cf.][]{Gronke2020Misty} as well as in the molecular case focused on here, the cloud seems to contract and then fragment upon expansion. \citet{McCourt2018} envisioned a `Jeans-like' instability in which hierarchical fragmentation occurs while the cloud cools, i.e., on a timescale of the (initial) cooling time. The fragmentation we observe occurs after expansion, that is, after the (much longer) sound crossing time of the cloud. This is closer to the picture conjectured by \citet{Waters2019} who studied thermal instability in 1D and found pulsations in the cold medium \citep[see also][]{Das2021} and extrapolated their findings to a regime where the pulsations are stronger than the restoring force leading to what they call `splattering'. 

In addition to this pulsational fragmentation, we see in our simulations fast rotations of the clumps -- induced by the contraction phase and the non-symmetric initial conditions -- which are the dominant process of further fragmentation. As rotation appears to offer a separate fragmentation mechanism than previously considered, we refer to this process as `\rattering.' It seems likely coagulation \citep[][]{waters2019coalescence,gronke2022cooling} averages out the spin, reducing further fragmentation. We plan to explore this in future work.

Even at large distances from galaxies, studies have shown that apart from the $\sim$10$^4\,$K `cold' phase even colder, molecular gas exists both in the CGM \citep[][]{Spilker2020} and ICM \citep[][]{jachym2019alma}. 
While in the CGM it is yet uncertain what the morphology of this molecular gas is, the existence of molecular filaments in cluster centers has been long known \citep[][]{salome2006cold}. 
However, there -- as well as in the CGM -- the origin of this molecular gas is unknown \citep[][]{Veilleux2020}.
In the context of the ICM, the $\sim$10$^4\,$K gas is thought to predominantly originate from thermal instability, i.e., to form in-situ \citep{Sharma2012,McCourt2012,DonahueVoit2022review} and potentially `shatter' during the cooling process \citep{McCourt2018}. 
Our results suggest that a similar process might be responsible to form a colder phase leading to a `fog' of $\lesssim 1000\,$K gas. 
For typical cluster conditions ($T\sim 10^{7-8}$\,K, $n\sim 0.1$\cc in the central regions to $n\sim 10^{-4}$\,\cc\ in the outskirts), the molecular shattering length-scale (cf. Eq.~\ref{eq:crit_ratio}) is  $\sim 0.01-100$\,pc. 
This would imply that fragments formed in such manner are very small, with a low volume filling fraction before they potentially coagulate to form the molecular streams we observe.
A similar `molecular fog' picture might apply to the CGM where our molecular length scale is (using $T_{\rm hot}\sim 10^6\,K$, $n_{\rm hot}\sim 10^{-2}$\cc) is $\sim$10\,pc.
That such a colder phase will form and molecules will form in it does naturally depend on the form of heating -- which is unknown in the ICM and CGM -- as well as on the dust abundance. Dust has been detected in both the CGM and ICM \citep[][]{ferrara1991evolution,elmegreen2000dust,Menard2010,PeekDUSTCIRCUMGALACTIC2015} but its origin and survival is theoretically not yet well understood \citep{rowlands2014dust,tumlinson2017circumgalactic}.

\vspace{-0.5cm}
\section{Conclusions}
\label{sec:Conclusions}
\vspace{-0.1cm}
We study the nonlinear fragmentation of rapidly cooling clumps of cold gas to molecular temperatures using three-dimensional inviscid hydrodynamic simulations. We find significant deviations in our results from previous work studying cooling down to only $10^4$\,K. Our specific conclusions are as follows:

\begin{enumerate}
    \item \textit{Size-Dependent Fragmentation Temperature}. For simulations that permit cooling below $10^4$\,K, sufficiently small clouds do not shatter, intermediate size clouds shatter but fail to cool below $10^4$\,K, while large clouds fragment and cool down to the temperature floor.
    \item \textit{Clumpy vs. Foggy}. As the size of the progenitor cloud increases, the number of clumps formed increases, the clumps form earlier, and the average rotational velocity of the clumps increases.
    \item \textit{Fragmentation Criterion}. We find fragmentation for clouds down to at least 400\,K follow the same analytical criterion min(c$_s$ t$_{\rm cool}$) with the minimum evaluated at the local minimum below (above) $\sim$8000\,K for molecular (atomic) shattering (cf. Eq.~\ref{eq:molshattering_crit}).
    \item \textit{Fragmentation Mechanism}. We find evidence suggesting rotation drives the fragmentation -- a process we dub \rattering.
\end{enumerate}

These results help to explain the cold gas structure in astrophysical conditions. However, the inclusion of additional physical effects such as magnetic fields, cosmic rays, conduction, viscosity, and chemical networks may introduce deviations from the picture outlined in this paper. We hope to explore the impact of these processes in future work.

\vspace{-0.6cm}
\section*{Acknowledgements}
\vspace{-0.1cm}
RJF gratefully acknowledges Syllvain Veilleux, Thorsten Naab, and Mateusz Ruszkowski for helpful discussions. 
MG thanks the Max Planck Society for support through the Max Planck Research Group.
The simulations and analysis in this work have been supported by the Max Planck Computing and Data Facility (MPCDF) computer clusters Cobra, Freya and Raven.
This project utilized the visualization and data analysis package \texttt{yt} \citep[][]{Turk2011}; we are grateful to the yt community for their support.

\vspace{-0.6cm}
\section*{Data Availability}
\vspace{-0.1cm}

Data related to this work will be shared on reasonable request to the corresponding author.


\vspace{-0.6cm}
\bibliographystyle{mnras}
\vspace{-0.1cm}
\bibliography{paper_molecular_shattering} 


\bsp	
\label{lastpage}
\end{document}